# Change in the Magnetic Domain Alignment Process at the Onset of a Frustrated Magnetic State in Ferrimagnetic La$_2$Ni(Ni$_{1/3}$Sb$_{2/3}$)O$_6$ Double Perovskite
## (Revised ..)

Diego G. Franco[1,2], Raúl E. Carbonio[1], and G. Nieva[2,3]

[1]INFIQC-CONICET, Depto. de Físico Química, Facultad de Ciencias Químicas, Universidad Nacional de Córdoba, Ciudad Universitaria. X5000HUA Córdoba, Argentina
[2]Laboratorio de Bajas Temperaturas. Centro Atómico Bariloche –CNEA. 8400 Bariloche, R. N., Argentina.
[3]Instituto Balseiro, CNEA and Universidad Nacional de Cuyo. 8400 Bariloche, R. N., Argentina.

**We have performed a combined study of magnetization hysteresis loops and time dependence of the magnetization in a broad temperature range for the ferrimagnetic La$_2$Ni(Ni$_{1/3}$Sb$_{2/3}$)O$_6$ double perovskite. This material has a ferrimagnetic order transition at ~100 K and at lower temperatures (~ 20 K) shows the signature of a frustrated state due to the presence of two competing magnetic exchange interactions. The temperature dependence of the coercive field shows an important upturn below the point where the frustrated state sets in. The use of the magnetization vs. applied magnetic field hysteresis data, together with the magnetization vs. time data provides a unique opportunity to distinguish between different scenarios for the low temperature regime. From our analysis, a strong domain wall pinning results the best scenario for the low temperature regime. For temperatures larger than 20K the adequate scenario seems to correspond to a weak domain wall pinning.**

## I. INTRODUCTION

MANY of the magnetic interactions found in transition metal oxide perovskites are due to superexchange and/or super-superexchange interactions mediated through the O$^{2-}$ *p* orbitals. In some materials the relative strength of these interactions determines the magnetic structure, range of the ordering temperatures and the possibility of frustration [1]-[5]. In the perovskite structure, the typical bond angles and distances usually favor antiferromagnetic superexchange interactions [6],[7]. However, in some special cases, due to disorder or differences in the magnetic state of the cations, a ferrimagnetic state is developed with macroscopic characteristics similar to a ferromagnetic state.

Coercivity and remanence are an indication of the metastability in ferromagnetic samples. Their magnitudes indicate how far the system is from equilibrium. They are related, therefore, with the relaxation to the equilibrium state, the anhysteretic [8] curve in the ferromagnetic state. In bulk ferromagnets the energy barriers that determine the time evolution of the magnetization are related to local interactions within a domain, the nucleation and the movement of domain walls (DW). The DW movement depends on the applied magnetic force, wall thickness and type and density of pinning centers.

In bulk ferromagnetic samples, a local frustration is normally hard to visualize due to the magnetic history dependence of the metastable states. However, the magnetic moments alignment within a domain and the movement of the DW have characteristic energies [9] that could be modified if some degree of magnetic frustration occurs at a microscopic level. This normally results in a strong DW pinning effect and causes an increase in the coercivity.

This article will present a detailed magnetic study of the ferrimagnetic double perovskite La$_2$Ni(Ni$_{1/3}$Sb$_{2/3}$)O$_6$. We will show that the material behaves as a ferrimagnet below 100 K and that there is a change in the magnetic domains alignment process at 20 K. We will show that below 20 K the hysteretic magnetic behavior is characteristic of a strong domain wall pinning regime due to the onset of a frustrated magnetic interaction.

## II. RESULTS

We prepared polycrystalline samples of La$_2$Ni(Ni$_{1/3}$Sb$_{2/3}$)O$_6$ by conventional solid-state reaction at 1400$^\circ$C [10]. X ray diffraction data from powders at room temperature showed the crystalline symmetry to be monoclinic, space group P2$_1$/n. This space group accommodates a rock salt arrangement of BO$_6$ and B'O$_6$ octahedra described by the a$^-$b$^-$c$^+$ system of three octahedral tilts in the Glazer's notation. The (Ni/Sb)$_{2d}$O$_6$ and (Ni/Sb)$_{2c}$O$_6$ octahedra are rotated in phase (along the primitive *c* axis) and out-of phase (along the primitive *a* and *b* axes). We performed a Rietvelt refinement of the structure using the FULLPROF program [11], resulting in lattice parameters of *a* = 5.6051(3) Å, *b* = 5.6362(3) Å, *c* = 7.9350(5) Å and $\beta$ = 89.986(4)$^\circ$. We refined the two crystallographic sites 2*d* and 2*c* with different occupancies Ni$^{2+}$/Sb$^{5+}$ to model the octahedral site disorder. The 2*d* cation site is almost fully occupied by Ni$^{2+}$ whereas the 2*c* site has occupancy close to 1/3 of Ni$^{2+}$ ions and 2/3 of Sb$^{5+}$. The resulting crystallographic formula can be written as La$_2$(Ni$_{0.976}$Sb$_{0.024}$)$_{2d}$(Ni$_{0.357}$Sb$_{0.643}$)$_{2c}$O$_6$.

The magnetic measurements were performed on polycrystalline pellets with a QD-MPMS SQUID magnetometer in the range 2 to 300K and -5 to 5T. In the main panel of Fig. 1 we show the magnetization, $M$, as a function of temperature, $T$, while cooling in a very low applied field, $H$. There is a transition to a magnetic polarized state at $T_C = 98(2)$ K.

We extrated the low temperature value of the saturation magnetization, $M_s$, from $M$ vs. $H$ curves, from the asymptotic extrapolation of the high field behavior with a Langevin function. This saturation magnetization, $M_s$, has a lower value than the one expected for the complete polarization of the $Ni^{2+}$ magnetic moments, 2.67 $\mu_B$/f.u.. Instead, the experimental $M_s$ value was 1.19 $\mu_B$/f.u., implying that the system behaves as a ferrimagnet, with two $Ni^{2+}$ magnetic sublattices antiferromagnetically coupled, one at the $2d$ site and another at the $2c$ site. The near 1/3 $Ni^{2+}$ random occupation of the $2c$ sites sublattice give as a result uncompensated $Ni^{2+}$ magnetic moments that order at 100 K. For a perfectly stoichiometric ferrimagnetic sample and full $Ni^{2+}$ occupancy of the $2d$ site $M_s$ should be 1.33 $\mu_B$/f.u., and lower values are expected if $Sb^{5+}$ partially occupies also the $2d$ site. The expected value for $M_s$ with the refined occupancies is 1.24 $\mu_B$/f.u., very close to the experimental one.

We measured hysteresis loops, $M$ vs $H$, for several temperatures below 100 K. We show in the inset of Fig. 1 a detail of the loops for 2 K and 20 K.

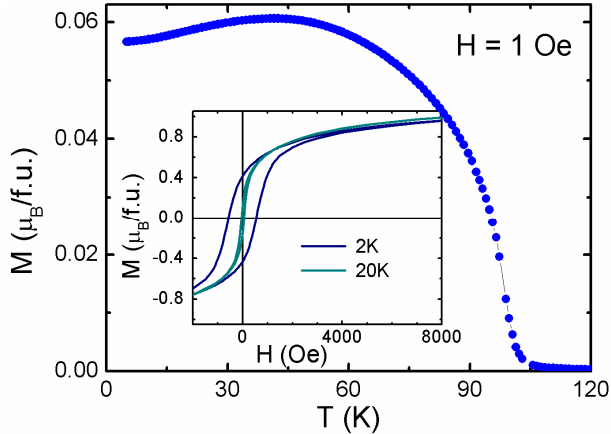

Fig. 1. (color online) Magnetization as a function of temperature cooling with an applied field of 1 Oe. Inset: Magnetization as a function of applied field, detail of the magnetization loops for two fixed temperatures T = 2 K and 20 K.

**FIG. 1 HERE**

We have also measured the time evolution of the magnetization at the coercive field (i.e. near the field for zero magnetization) after saturation at 1, 3 and 5 T for each temperature. We show in Fig. 2 typical $M$ vs time data, for three different temperatures at their corresponding coercive fields.

**FIG. 2 HERE**

### III. DISCUSION

We show in Fig. 3(a) and (b) the temperature dependence of the coercive field, $H_c$, and the ratio between remanent magnetization and saturation magnetization, $(M_r / M_s)$. The general feature observed in Fig. 3 is that $M_r$ and $H_c$ increase steeply when the temperature is lowered below 20 K indicating an increase in the energy absorbed by the material to change the direction of $M$.

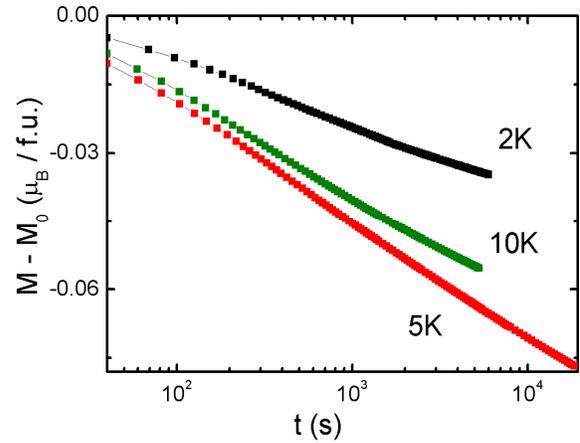

Fig. 2. (color online) Difference between the measured magnetization and the initial one, $M_0$, as a function of time. The shown magnetization time dependence was taken at $H_c$ ($M \sim 0$) after saturation in the opposite direction. We indicate the fixed temperatures for each experiment.

The measured values of the coercive field, $H_c$, display two different regimes as can be seen in Fig. 3(a). For $T > 20$ K a linear behavior of $H_c$ was found. This linear behavior is characteristic of weak DW pinning (WDWP), produced by a random distribution of individual weak pinning sites [9]. In this case the coercive field is given by

$$H_c = H_{0W}\left[1 - \left(\frac{25 k_B T}{31 \gamma b^2}\right)\right] \qquad (1)$$

where $H_{0W}$ is the zero temperature extrapolated reversion field, $k_B$ is the Boltzmann constant, $\gamma$ is the DW energy per unit area and $b$ is a measure of the DW thickness. The obtained values are shown in Table I.

In the low temperature regime, $T < 20$ K, two models describe reasonably well the data. One corresponds to strong DW pinning (SDWP),

$$H_c = H_{0S}\left[1 - \left(\frac{75 k_B T}{4bf}\right)^{2/3}\right]^2 \qquad (2)$$

where $H_{0S}$ is the coercive field at zero temperature and $f$ is the magnetic force needed to depin a domain wall. The fitted values are shown in Table I.

The other model corresponds to the freezing of single domain large particles (SDLP) or clusters [11], [12]. In this scenario,

$$H_c = H_K\left[1 - \left(\frac{25 k_B T}{KV}\right)^{1/2}\right] \qquad (3)$$

where $H_K$ is the anisotropy field of a particle or cluster, $V$ is its volume and $K$ is the uniaxial anisotropy energy density. The fitted values are shown in Table I.

**FIG. 3 HERE**

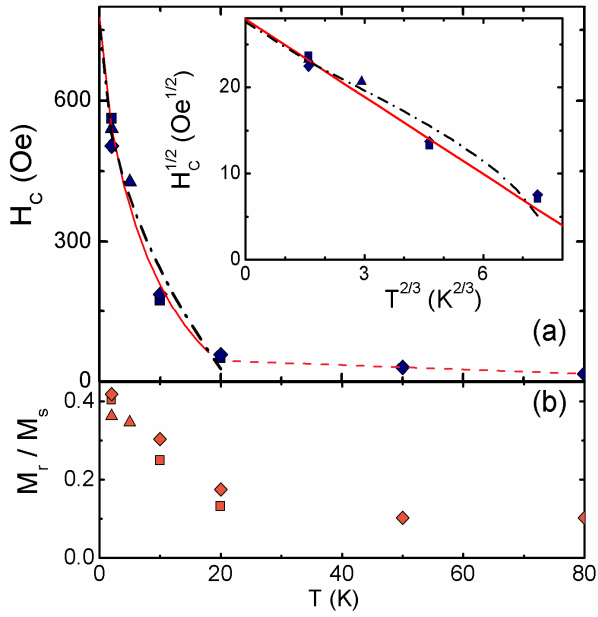

Fig. 3. (color online) (a) Coercive field $H_c$ vs $T$ for $La_2Ni(Ni_{1/3}Sb_{2/3})O_6$ polycrystalline samples. The solid lines (SDWP model), dash-dotted lines (SDLP freezing model) at low temperature and the dashed line (WDWP model) at high temperature are fits described in the text. Inset: $H_c^{1/2}$ vs $T^{2/3}$ showing the linear behavior expected in the SDWP model, solid lines are the SDWP model and dash-dotted lines are the SDLP freezing model. (b) Normalized remanent magnetization (at $H = 0$) vs $T$. The different symbols indicate different samples in both panels.

In Fig. 3(a) and the inset we show the lines (solid and dash doted) corresponding to each model. The best fit is obtained with the SDWP model but the freezing of SDLP model is also in fair agreement with the $H_c$ data.

The time evolution of the magnetization could be used to discern between the two scenarios at low temperature. If a distribution of activation energies is present in the material [9] a logarithmic behavior is expected for $M(t)$:

$$M = M_0 - S \ln(t) \quad (4)$$

where $M_0$ is the starting value of the magnetization and $S$ the magnetic viscosity coefficient. The above relation holds approximately for our polycrystalline pellets samples as we show in Fig. 2.

In a tipical ferromagnet, the time dependence of $M$ is irreversible and this behavior has been connected with the irreversibility caused by a small change in field, the so called irreversible susceptibility, $\chi_{irr}$. Both irreversibilities are related by a fictitious field, the fluctuation field, $H_f$, in the theory introduced by Neel [13] that represents an average of the thermally activated, time dependent processes [14] leading to equilibrium by reversing the metastable magnetization. In terms of the magnetization derivatives, at a given field and temperature,

$$S = \left(\frac{\partial M_{irr}}{\partial \ln(t)}\right)_{H_i} = \left(\frac{\partial M_{irr}}{\partial H_i}\right)_t \left(\frac{\partial H_i}{\partial \ln(t)}\right)_{M_{irr}} = \chi_{irr} H_f \quad (5)$$

where $M_{irr}$ is the irreversible magnetization and $H_i$ is the internal field. In the case of a time independent viscosity coeficient $S$, the fluctuation field is equivalent to the magnetic viscosity parameter $S_v$, that can be written in terms of the activation energy, $E$, necessary for magnetization reversal [15],

$$S_v = \frac{-k_B T}{(dE/dH)_{M_{irr}}} = \frac{S}{\chi_{irr}}. \quad (6)$$

To determine the temperature dependence of $S_v$, a value of $H$ equal to the coercive field is chosen [15] ($M_{irr} = 0$). The activation energy for the SDWP, WDWP and clusters or SDLP freezing are given by [16]:

$$E_S = (4bf/3)\left[1 - \left(\frac{H}{H_{0S}}\right)^{1/2}\right]^{3/2} \quad (7)$$

$$E_W = (31\gamma b^2)\left[1 - \left(\frac{H}{H_{0W}}\right)\right] \quad (8)$$

$$E_F = KV\left[1 - \left(\frac{H}{H_K}\right)\right]^2 \quad (9)$$

and in each case the magnetic viscosity parameters are given respectively by,

$$(S_v)_S = \frac{4}{75} H_{0S}\left(\frac{75 k_B T}{4bf}\right)^{2/3}\left[1 - \left(\frac{75 k_B T}{4bf}\right)^{2/3}\right] \quad (10)$$

$$(S_v)_W = \frac{1}{25} H_{0W} \frac{25 k_B T}{31\gamma b^2} \quad (11)$$

$$(S_v)_F = \frac{1}{50} H_K \left(\frac{25 k_B T}{KV}\right)^{1/2}. \quad (12)$$

From the experimental data (such as those of the inset of Fig. 4) the values of $\chi_{irr}(H = H_c)$ can be extracted. They can be approximated as those of the total $\chi$ at $H_c$, neglecting the reversible contribution to $\chi$ [14]. These values are shown in Fig. 4. Also from the experimental data the $S$ values can be calculated from (4) at $H_c$, since the linear behavior holds, Fig. 2. In this case we take $M_{irr}$ as the measured $M$, neglecting the reversible component.

**FIG. 4 HERE**
**TABLE 1 HERE**

The experimental values of $S_v$ were obtained by using (6). They are displayed in Fig. 5 together with the fits for different models in different temperature ranges using (10) - (12). At low temperature, the best fit to the data is given

TABLE I

|  | $\gamma b^2$ ($10^{-14}$ erg) | $H_{0W}$ (Oe) | $4bf$ ($10^{-13}$ erg) | $H_{0S}$ (Oe) | $KV$ ($10^{-14}$ erg) | $H_{0F}$ (Oe) |
|---|---|---|---|---|---|---|
| $H_c$ | 1.26 | 53.5 | 3.07 | 780 | 7.4 | 760 |
| $S_v$ | 1.4 | 53.5 | 2.13 | 780 | 7.4 | 760 |

Fitted parameters for the WDWP model above 20 K and the two models compared below 20K, SDWP and freezing of SDLP. The first column indicates whether the coercive field or the magnetic viscosity parameter were used in the parameters determination. In the case of the viscosity parameter, the zero temperature fields $H_{0W}$, $H_{0S}$ and $H_{0F}$ were not fitted but taken from the $H_c$ fits.

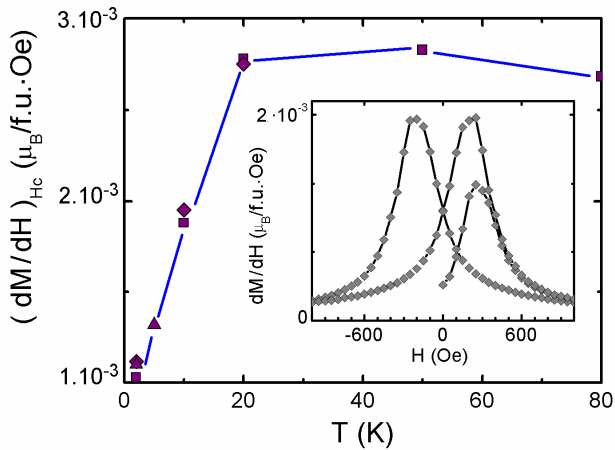

Fig. 4. (color online) Total susceptibility $\chi = dM/dH$ measured at $H_c$ vs $T$ for the polycrystalline pellets. Different symbols indicate different samples. The line is a guide to the eye. The inset shows a typical $\chi$ vs $H$ curve taken at 10 K.

using (10) (a SDWP scenario) provided the nonmonotonic behavior of $S_v$. The parameters obtained are shown in Table I. Clearly no good agreement is found for the freezing of clusters or SDLP scenario. In the high temperature region the experimental $S_v$ vs $T$ is in agreement with the linear behavior calculated in (11). However, a non-zero $S_v$ ($T = 0$) value was found, not present in the model.

**FIG. 5 HERE**

Therefore, based on combined data extracted from the hysteresis loops and time dependence of the magnetization, we depicted two regimes in La$_2$Ni(Ni$_{1/3}$Sb$_{2/3}$)O$_6$ pelletized polycrystalline samples: Weak pinning of DWs at $T > 20$ K and strong pinning of DWs below that temperature. The microscopic origin of this change of regime could be related with the onset of a super-superexchange antiferromagnetic interaction among Ni$^{2+}$ via O$^{2-}$ -Sb$^{5+}$ - O$^{2-}$ paths [10] that creates a frustrated magnetic interaction.

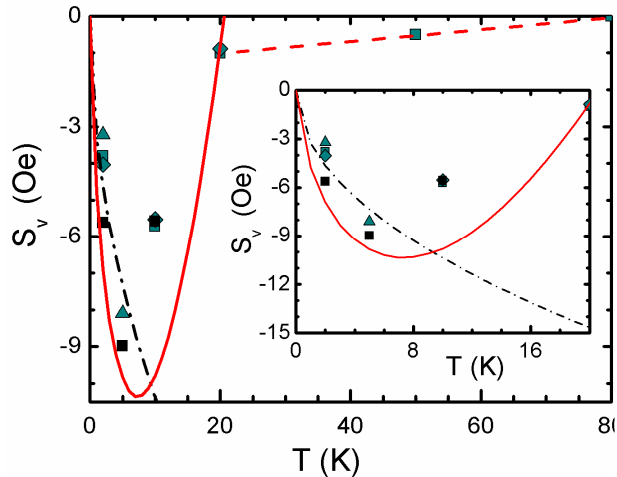

Fig. 5. (color online) Magnetic viscosity parameter $S_v$ vs $T$ for La$_2$Ni(Ni$_{1/3}$Sb$_{2/3}$)O$_6$ polycrystalline pellets. Different symbols indicate different samples, the solid symbols corresponds to $S_v$ values calculated with the data displayed in Fig.2. The lines represent the models described in the text, SDWP model (solid line), freezing of SDLP model (dash dotted line) and WDWP model (dash line). In the inset, a zoom of the low temperature region is shown.

## IV. CONCLUSION

The ferrimagnetic state in La$_2$Ni(Ni$_{1/3}$Sb$_{2/3}$)O$_6$ was found to be characterized by two different regimes for domain wall movement, a strong and a weak domain wall pinning regime at low and high temperatures respectively. The temperature range of the strong domain wall pinning regime coincides with that of the existence of a proposed frustrated state. The scenario of clusters or large single domain particles freezing was discarded based in the coercive field and magnetic viscosity parameter temperature dependence analysis.

### ACKNOWLEDGMENT

We thank E.E. Kaul for fruitfull discussions. R.E.C, and G.N. are members of CONICET. D.G.F. has CONICET scholarship. Work partially supported by ANPCyT PICT07-819, CONICET PIP 11220090100448 and SeCTyP-UNCuyo 06/C313. R.E.C. thanks FONCYT (PICT2007-303), CONICET (PIP 11220090100995) and SECYT-UNC (Res. 214/10) for finantial support.